\newcommand{\beq}{\begin{equation}}
\newcommand{\eeq}{\end{equation}}
\newcommand{\bea}{\begin{eqnarray}}
\newcommand{\eea}{\end{eqnarray}}

\newcommand{\oncite}{\onlinecite}
\newcommand{\bi}{\bibitem}
\newcommand{\vareps}{\varepsilon}
\documentstyle[aps,prb,multicol,graphicx]{revtex}
\begin{document}
\draft
\title{Magnetic-field--induced Luttinger liquid}
\author{C.\ Biagini$^{1,2}$, D.\ L.\ Maslov$^{2}$,
M.\ Yu.\ Reizer$^{2}$\cite{reyzer}, and L.\ I.\ Glazman$^{3}$}
\address{
$^{1)}$Istituto Nazionale di Fisica della Materia,
L.go E. Fermi 2, 50125, Firenze, Italy\\
$^{2)}$Department of Physics, University of
Florida, P.\ O.\ Box 118440, Gainesville, Florida 32611-8440\\
$^{3)}$University of Minnesota, Theoretical Physics Institute,
116 Church Street, SE Minneapolis, MN 55455}
\maketitle \centerline{(\today)}
\begin{abstract}
It is shown that a strong magnetic field applied to a bulk metal
induces a Luttinger-liquid phase. This phase is characterized
by the zero-bias anomaly in tunneling: 
the tunneling conductance scales as a power-law of voltage
or temperature. The tunneling exponent increases with the magnetic
field as $B\ln B$. The zero-bias anomaly is most pronounced
for tunneling with the field applied perpendicular to the plane of the
tunneling junction.
\end{abstract}
\begin{multicols}{2}
A strong magnetic field applied to a bulk metal tends to reduce the effective
dimensionality of charge carriers from 3D to 1D. This feature is
most pronounced in the ultra-quantum limit, when only the lowest
Landau level remains populated. The reduction of the effective dimensionality
in a system of interacting particles 
is expected to result in a number of unusual phases, which
are one-dimensional in nature, such as spin-and charge-density waves (CDW)
\cite{celli-mermin,fukuyama,yakovenko}, Wigner crystal \cite{wc}, excitonic insulator (EI)
\cite{abrikosov},
re-entrant superconductor \cite{tesanovic}, etc.
A field-induced CDW is believed to have been observed in graphite
(see, e.g., 
Ref.~[\oncite{shimamoto}] for an extensive
bibliography and discussion). 
There have been also earlier reports of tentative field-induced EI 
transitions in Bi$_{1-x}$Sb$_x$ \cite{brandt} and Bi \cite{mase}.

In this paper, we focus on another field-induced state, which is not
related to any instability, but evolves adiabatically from the
conventional three-dimensional (3D) Fermi-liquid as the magnetic
field increases. This state is a Luttinger Liquid (LL), whose
existence can be anticipated from the following simplified picture. In
a strong magnetic field, electron trajectories are helices spiraling
around the field lines. A bundle of such trajectories with a common
center of orbit can be viewed as a 1D conductor (\lq\lq wire\rq\rq\/)
with the Fermi velocity $v_F^B$ determined from the condition of the
fixed carrier density $n$ in the bulk, 
\beq
v_F^B=2\pi^2n\ell_B^2/m_{||}.
\label{partition}
\eeq 
Here $\ell_B=(eB)^{-1/2}$ is the magnetic length, and $m_{||}$ is
the effective mass along the field; we assume also full spin
polarization of electrons, and use the units with $\hbar\!=\!c\!=\!1$.
In the presence of electron-electron interactions, each ``wire'',
considered separately, is in the LL-state.  Interactions with
small-momentum transfers among electrons on different ``wires'' do not
change the LL-nature of a single-wire state \cite{DL}.

The quantity of primary interest of this paper is the experimentally
measurable density of states (DOS) at the sample boundary. We
calculate this quantity first by treating perturbatively the
interaction between electrons moving in 3D
space in the presence of a quantizing magnetic field. We demonstrate 
that the boundary DOS exhibits  a characteristic  for a
Luttinger liquid \cite{kf} power-law anomaly
at the Fermi level:
\beq
\nu(\varepsilon)\propto|\varepsilon-\vareps_F^B|^{\alpha_s},
 \label{nubulk}
\eeq
where $\vareps_F^B=m_{||}(v_F^B)^2/2$ is the Fermi energy of the 1D motion
along the magnetic field.

We establish the correspondence between 
a 3D electron plasma in a
quantizing magnetic field and a Luttinger liquid 
by considering the
electron-electron interaction in the basis of coherent states (CS) on
the von Neumann lattice\cite{note}. In this basis, a system of 3D electrons in
the ultra-quantum limit is equivalent to a lattice of 1D wires
pointing in the direction of the field and separated by distance
$\sqrt{2\pi}\ell_B$, similar to the simplified picture mentioned
above \cite{note2}.  The problem of an array of wires coupled via the long-range
Coulomb interaction [Dzyaloshinskii-Larkin (DL) model \cite{DL}] can
be solved exactly either via summation of diagram series \cite{DL} or
bosonization \cite{schulz}. Utilizing the latter technique, we derive
the ``bulk'' tunneling density of states.
This quantity also exhibits a power-law anomaly (\ref{nubulk})
with a different exponent $\alpha$. Different values
of bulk and boundary exponents is another 
characteristic feature of a LL \cite{kf}.
We calculate explicitly the values of the exponents $\alpha$ and
$\alpha_s$ in the limit of a long-range interaction potential,
\beq
\kappa^B\ell_B\ll 1,
\label{condition} 
\eeq
where $\kappa^B=\omega_p/v_F^B$ is the (field-dependent) inverse
screening length, and $\omega_p$ is the zero-field plasma frequency.
Condition (\ref{condition}) is satisfied in magnetic fields
$B\lesssim B_0\equiv\pi^3 \epsilon n/m_{||}e^3$, where $\epsilon$ is
the background dielectric constant. The introduced here characteristic
magnetic field $B_0$ can be expressed as $B_0\simeq B_q/r_s$ in terms
of the magnetic field $B_q$ de-populating all but the lowest Landau
level, and of the gas parameter $r_s$. Under
condition (\ref{condition}), we find
\beq 
\alpha_s=2\alpha\ln\frac{1}{\sqrt{\alpha}}, \quad
\alpha=\frac{e^2}{2\pi\epsilon v_F^B}\equiv\frac{B}{4B_0}.
\label{alphas}
\eeq
The tunneling current-voltage characteristic is thus a power-law 
\beq
I\propto |V|^{\alpha_s+1}\rm{sgn}V
\label{cvc}
\eeq
for $e|V|\ll\vareps_F^B$.
Eq.~(\ref{alphas}) refers to the situation when the magnetic field
is perpendicular to the plane of the tunneling contact. If the
field is parallel to the contact plane, electrons move along skipping
orbits. Interactions between electrons moving in the same
direction do not result in anomalous scaling, therefore there
is no tunneling anomaly for this field orientation. This means
that by tilting the field one can vary the tunneling exponent from
its maximum value, given by Eq.~(\ref{alphas}), to zero.

Another indication of the tunneling anomaly comes from the gapless
behavior of a plasmon in a strong magnetic field. For
$\omega_c\!\gg\!\omega$, where $\omega_c=eB/m_c$ is the cyclotron
frequency, the dispersion relation for a classical magnetoplasmon is
given by $\omega ^{2}\!=\!\left(k_z^2/k^2\right)\!\omega _{p}^{2}$
[\oncite{physkin}].  As we see, $\omega\!\to\! 0$ for $k_z\!\to\! 0$
at finite $k$.  A gapless charge mode slows down the relaxation of an
excessive charge added to the conductor in a tunneling event. This
mechanism is responsible for the zero-bias anomalies in tunneling into
disordered metals and 1D Luttinger liquids.  A metal in a strong
magnetic field provides one more example of such a behavior.

The LL is destroyed either by backscattering between electrons on
different \lq\lq wires\rq\rq\/, which results in a CDW-instability, or
by formation of bound electron-hole pairs (for the case of a
semimetal), which leads to an EI-instability. Also, impurity
scattering transfers electrons between the \lq\lq wires\rq\rq\/ and
thus destroys the 1D motion. The relevant energy scales for these
processes is the energy gap, $\Delta$, of either CDW- or EI-origin,
and the level broadening, $\Gamma$, respectively. On the other hand, a
LL-behavior sets in at energies smaller than the Fermi energy
$\vareps_F^B$. The gap $\Delta$ increases with the field, whereas
$\vareps_F^B$ decreases with the field. The LL-state should thus exist in
the energy interval which narrows down as $B$ increases: 
\beq
\rm{max}\{\Delta(B),\Gamma\}\lesssim\vareps\ll\vareps_F^B\propto
B^{-2}.
\label{condition2}\eeq

Before
giving
the derivation of the results announced above, we would like
to discuss their relevance to the experiment. The search for LL-like
tunneling anomalies in truly 1D systems (edges
of 2D electron gas in the FQHE regime \cite{chang}, carbon nanotubes
\cite{nanotube}, and quantum wires\cite{wires}) 
has been a very active but remarkably difficult area over the last few years.
Compared to experiments on 
tunneling into 1D systems, tunneling into
a 3D system in the ultra-quantum limit has the obvious advantage of
the macroscopic system size. It is also advantageous that the tunneling exponent
is a function of the external parameters (magnitude and
direction of the magnetic field), which can be varied over a wide range.

The choice of the right material is crucial. Among the low-density
semimetals (Bi and its alloys and graphite) used conventionally for
high-field studies, graphite seems to be the optimal candidate.  For
$B$ along the c-axis, the ultra-quantum limit is achieved at
$B\!\simeq\!7$\, T. With material parameters for
graphite\cite{brandtbook} ($m_{||}\!=\!10 m_0$, $\epsilon\!=\!6$, and
$n\!=\!3\times 10^{18}$\,cm$^{-3}$), Eq.~(\ref{alphas}) gives
$\alpha\!=\!0.8$ already at 7\,T.  For comparison, an anomalously
large value of the background dielectric constant ($\epsilon\simeq
100$) for Bi \cite{edelman} and a smaller value of $m_{||}$ ($=0.5m_0$
for holes)  
makes $\alpha$ to be very small: even
for $B\!=\!100$\,T, one has $\alpha\!=\!0.05$.  According to a
recent experiment \cite{singleton}, the field-induced CDW gap in
graphite $\Delta\!\lesssim\!1$\,K for $B\!\lesssim\!30$\,T and takes
its maximum value ($\Delta\!\simeq\!10$\,K) at $B\!\simeq\!50$\,T.
For fields in the range $\!7\,{\rm T}\lesssim\!B\!\lesssim\!30$\,T,
inequality (\ref{condition2}) is satisfied for more than two decades
of energies (voltages).

In the limit of weak electron-electron interactions, one can derive
the field-induced zero-bias tunneling anomaly (\ref{cvc}) by
calculating interaction corrections to the transmission coefficient
through a tunneling barrier \cite{yue}.  These corrections can be then
summed up by using the renormalization group procedure.

Let a potential barrier of transmission coefficient $t_{0}\!\ll\!1$
separate two metallic half-spaces, $z\!<\!0$ and $z\!>\!0$. The
magnetic field is perpendicular to the contact plane. We assume that
the base-electrode ($z\!<\!0$) is made of a high-density metal and
therefore is not affected by the magnetic field. Electron-electron
interactions are then not important in this half-space and we treat it
as a Fermi gas. The counter-electrode ($z\!>\!0$) is a low-density
metal in the ultra-quantum magnetic limit.  In this half-space,
electrons interact via potential $U({\bf r}-{\bf r}')$. Keeping only
the leading terms in $t_{0}\!\ll\!1$, the free wave-function in the Landau
gauge  ${\bf A}\!=\!(-yB,0,0)$ for
$z\!>\!0$ and far away from the barrier is given by
\begin{mathletters}
\begin{eqnarray}
\Psi _{p_{x},-p_{z}}^{0} &=&-\left( 2i\sin p_{z}z/\sqrt{L_{z}}\right) \Phi
_{p_{x}}^{0}({\bf r}_{\perp })  \label{psimin} \\
\Psi _{p_{x},p_{z}}^{0} &=&t_{0}\left( e^{ip_{z}z}/\sqrt{L_{z}}\right) \Phi
_{p_{x}}^{0}({\bf r}_{\perp }),  \label{psimax}
\end{eqnarray}
\end{mathletters}
where  $p_{z}\!=\!\sqrt{2m\varepsilon} >0$, ${\bf r}_{\perp}\equiv(x,y)$ and
\[
\Phi_{p_{x}}^{0}({\bf r}_{\perp})=(\sqrt{\pi}\ell_{B}L_x)^{-1/2}
e^{ip_{x}x}\exp \left[ -(y+p_{x}\ell _{B}^{2})^{2}/2\ell _{B}^{2}\right].
\]
The leading correction to the transmission coefficient comes from the
exchange interaction\cite{yue}, if the interaction potential is smooth
on the scale of Fermi wavelength, see Eq.~(\ref{condition}).

The correction to the wave caused by the exchange potential $V_x$ has
the form 
\bea
\Psi _{p_{x},p_{z}}({\bf r})&=&\Psi_{p_{x},p_{z}}^{0}({\bf r})+\int\int
d{\bf r}'d{\bf r}''G_{\varepsilon_{z}}^{>}({\bf r},
{\bf r}')\nonumber\\
&&\times V_{x}({\bf r}',{\bf r}'')\Psi
_{p_{x},p_{z}}^{0}({\bf r}'')\bf{,\ }
\label{inteq}\eea
where $G_{\varepsilon_{z}}^{>}({\bf r},{\bf r}')$ is the Green's function
of the free Schr\"odinger equation in the half-space $z\!>\!0$, and
$\varepsilon_z\equiv p_z/2m^2$ is the energy of electron motion along
the field. The integration in (\ref{inteq}) goes only over those regions
where electrons interact, i.e., $z',z''\!\geq\!0$.

The renormalized transmission coefficient is extracted from the
asymptotic form of $\Psi_{p_{x},p_{z}}$ for $z\to+\infty$ which, in its
turn, is determined by the asymptotic behavior of
$G_{\varepsilon_{z}}^{>}({\bf r},{\bf r}')$ [see
Eq.~(\ref{inteq})]. The asymptotic form of the Green's function 
in the
infinite
space can be written as $G_{p_{z}}({\bf r},{\bf r}')|_{z\to
+\infty}\!=\!G_{||}(z,z',p_{z})G_{\perp }({\bf r}_{\perp },{\bf
r}_{\perp }'), $ where $
G^{||}_{\varepsilon_z}(z,z')\!=\!(m_{||}/ip_{z})e^{ip_{z}|z-z'|}$ is the Green's
function for the 1D motion along the $z$-axis, and
\begin{equation}
G_{\perp }({\bf r}_{\perp },{\bf r}_{\perp }')=\!\frac{1}{2\pi \ell _{B}^{2}}
e^{-\left[ {\bf r}_{\perp }-{\bf r}_{\perp }'%
\right]^{2}/4\ell _{B}^{2}}e^{-i(y+y')(x-x')/2\ell _{
B}^{2}}.
\label{gperp}\end{equation}
The asymptotic form of $G_{\varepsilon_{z}}^{>}({\bf r},{\bf r}')$ 
is then  obtained via the
method of images 
\[
G_{\varepsilon_{z}}^{>}({\bf r},{\bf r}')|_{z\to +\infty}=
\left[ G^{||}_{p_{z}}(z,z')-G^{||}_{p_{z}}(z,-z')\right] 
G_{\perp }({\bf r}_{\perp },%
{\bf r}_{\perp }').
\]
The exchange potential is given by
\bea
V_{x}({\bf r,r}')&=&-U({\bf r-r}')\int_{0}^{p_{F}^B}\frac{%
dp_{z}}{2\pi }\int_{-\infty }^{\infty }\frac{dp_{x}}{2\pi }\left[ \Psi
_{p_{x},-p_{z}}^{0}({\bf r}')\right] ^{\ast }\nonumber\\
&&\times\Psi_{p_{x},-p_{z}}^{0}({\bf r})
\nonumber\\
&&\approx-U({\bf r-r}')\frac{\sin
p_{F}^B(z+z')}{\pi(z+z')}G_{\perp }({\bf r}_{\perp },{\bf r}_{\perp }'),
\label{vx}\eea
where $p_{F}^B=m_{||}v_F^B$.  In Eq.~(\ref{vx}), we retained only that
term which at $p_z\approx p_F^B$ leads to a logarithmic divergence in
the integral over $z+z'$ in Eq.~(\ref{inteq}).

The transmission
coefficient in the first order with respect to interactions
is
\beq
t=t_0\left(1-\frac{e^2}{2\pi \epsilon v_F^B}\ln \frac1{\kappa ^B\ell _B}
\ln\frac1{\ell _B|p_z-p_F^B|}\right),
\label{transm}\eeq
where the inverse screening radius
\beq
\kappa ^{B}=\omega _{p}/v_{F}^{B}=
2\alpha{\ell_B}^{-1}\propto B.
\label{kappaB}
\eeq
This result is valid if both logarithmic factors are large. The
contribution $t$ from the Hartree term is smaller than (\ref{transm})
by $\ln (1/\kappa ^{B}\ell _{B})$. The higher order terms in the
expansion of $t$ can be summed up via the renormalization
group procedure \cite{yue}.  The result is
\begin{equation}
t
\propto |p_{z}-p_{F}^B|^{\alpha_s/2}
\propto |\varepsilon -\varepsilon_{F}^{B}|^{\alpha_s/2}.
\label{rg}
\end{equation}
As $I\propto |t|^{2}V$, Eq.~(\ref{rg}) yields Eq.~(\ref{cvc}) with
exponent (\ref{alphas}). The perturbative result (\ref{rg})
is valid for $\alpha\ll 1$, which means that $\alpha_s$ is also small within
this approach.

We now demonstrate how the LL-state emerges in the coherent state formulation.
The method of bosonization, applied 
in this formulation, provides a simple tool for calculation the ``bulk'' 1D
density of states\cite{kf}. Also, it allows us to estimate the
tunneling exponents for stronger interaction ($\alpha\gtrsim 1$).
The $\Psi$-operator is expanded over the coherent states basis 
\begin{equation}
\Psi ({\bf r})=\sum_{N}\sum_{{\bf R}}a_{N{\bf R}}^{\dagger
}(z)\chi _{N{\bf R}}({\bf r}_{\perp }),
\label{expansion}
\end{equation}
where ${\bf r}\!\!=\!\!(z,{\bf r}_{\perp }), {\bf R}\!\!=\!\!(R_{x},R_{y})$.
A coherent state $\chi _{N{\bf R}}({\bf r}_{\perp })$ is a simultaneous
eigenstate of the energy operator 
\[
(2m_{c})^{-1}\left({\bf p}_{\perp}-
e{\bf A}\right)^{2}
\chi_{N{\bf R}}=\omega _{c}(N+1/2)\chi_{N{\bf R}},
\]
and of the ``guiding center'' operator 
\[
(\hat{x}_{0}-i\text{ }\hat{y}_{0}) \chi_{N{\bf R}}=(R_{x}-iR_{y})\chi
_{N{\bf R}},
\]
where
$\hat{x}_{0}\!=\!\hat{x}\!+\!(m\omega _c)^{-1}(\hat{p}_{y}-eA_y)$ and
$\hat{y}_{0}\!=\!\hat{y}\!-\!(m\omega _c)^{-1}(\hat{p}_{x}-eA_x)$
are the operators corresponding to the classical coordinates of guiding
centers \cite{malkin}. In the symmetric gauge, ${\bf A}=(1/2){\bf B}\times {\bf r}$, an
explicit form of the CS corresponding to $N=0$ is \cite{malkin}
\[
\chi _{0{\bf R}}({\bf r}_{\perp })=\frac{1}{\sqrt{2\pi }\ell _{B}} \exp \left[ -\frac{
\left( 
{\bf r}_{\perp }-{\bf R}
\right)
^{2}+2i{\bf r}_{\perp }\wedge {\bf R}
}
{4\ell _{B}^{2}}
\right],
\]
where ${\bf a}\wedge{\bf b}=a_xb_y-a_yb_x$.

As any set of coherent states, the set of $\chi _{N{\bf R}}$ 
remains (over) complete even if
defined on a subset of points in the complex plane, $R_{x}+iR_{y}$,
rather than on the whole plane. The 
von Neumann theorem \cite{neumann}
allows one to choose this subset as
sites of a 2D square lattice.
To fix the value of the lattice spacing, one notices that
a total of $n$ particles per unit volume are now distributed over ``wires'',
which are parallel to the magnetic field and cross the plane perpendicular
to the field at the sites of the von Neumann lattice.
For $N\!=\!0$, the number density per unit length of
each ``wire'' is $p_{F}^{B}/\pi $, their
the areal density is $1/s^{2}$, thus
$n\!=\!p_{F}^{B}/\pi s^{2}$.
Comparing this equation to (\ref{partition}),
one gets $s=\!\sqrt{2\pi }\ell _{B}$. 
This is a maximum value of the lattice spacing for
which the set is still complete.
States of the CS basis on the von Neumann lattice describe particles which are
localized on the lattice sites in the direction transverse to the field but
free to move along the field. This picture is in a close resemblance to the
classical trajectories spiraling around the field lines.

Coherent states are not orthogonal. In particular, 
\beq
\langle\chi_{0{\bf R}}|\chi_{0{\bf R}'}\rangle=
\frac{1}{2} \exp \left[ -\frac{
\left( 
{\bf R}-{\bf R}'
\right)
^{2}+2i{\bf R}\wedge {\bf R}'
}
{4\ell _{B}^{2}}
\right],
\eeq
which can be replaced by $2\pi\ell_B^2\delta^2({\bf R}-{\bf R}')$ in the limit
$\ell_B\to 0$. Only in this limit the coherent
set diagonalizes the quadratic part of the Hamiltonian.
Physically, this limit means that $\ell_B$ is much
less than the characteristic value of $|{\bf R}-{\bf R}'|$.
For interacting electrons, 
typical $|{\bf R}-{\bf R}'|$ is of the order of
the 
screening radius $1/\kappa^B$.
Thus the approach based on coherent states
is valid for a long-range interaction as defined by condition (\ref{condition}).

Using asymptotic
orthogonality of the coherent states, one arrives at the effective 
1D Hamiltonian (from now on we concentrate on $N=0$ and suppress 
index $N$)
\bea
  \label{1d.1}
        H&=&\sum_{\mathbf R}
        \int dza^\dagger_{\mathbf R}(z)\left[
  \frac{\omega_c}{2}+   \frac{p_z^2}{2m_{||}}\right]a_{\mathbf R}(z)\nonumber\\
&+&
        \frac{e^2}{2\epsilon}\sum_{{\mathbf R},{\mathbf R}^\prime}
        \int dzdz^\prime \frac{a^\dagger_{{\mathbf R}}
        (z)a^{\dagger}_{{\mathbf R}'}(z')\hat 
a_{{\mathbf R'}^\prime}
        (z^\prime)a_{{\mathbf R}}(z)}{\sqrt{\left({\mathbf R}-{\mathbf R}^\prime\right)^2+
        \left(z-z^\prime\right)^2}}.    
\label{ham1} \eea
Due to the long-range nature of the Coulomb potential, it suffices
to keep only the forward scattering processes in (\ref{ham1})
(both for ${\bf R}={\bf R}'$ and ${\bf R}\neq{\bf R}'$), in which case
 (\ref{ham1}) is identical to the DL model \cite{DL}.
Backscattering for ${\bf R}\neq{\bf R}'$ leads to a CDW instability
\cite{fukuyama,babichenko}. The forward scattering part
of (\ref{ham1}) is diagonalized via bosonizing the fermions
\[
a_{{\bf R}}=:e^{-i\left(p_{F}^{B}z+\sqrt{\pi }\{\phi
{{\bf R}}+\theta _{{\bf R}}\}\right)} +e^{i\left(p_{F}^{B}z+\sqrt{\pi }%
\{\phi _{{\bf R}}-\theta _{{\bf R}}\}\right)} :,
\]
where $[\phi_{{\bf R}}(z_1),\partial_{z_2}\theta_{{\bf R'}}
(z_2)]=i\delta_{{\bf R},{\bf R}'}\delta(z_1-z_2)$. 
The result of a straightforward calculation 
for the equal-point  (Matsubara) Green's function is
\begin{mathletters}
\begin{eqnarray}
{\cal G}(\tau ) &\sim &\tau ^{-1}e^{-f(\tau )},\nonumber\\ 
f(\tau ) &=&2\pi ^{2}\ell _{B}^{2}\!\int_{{\rm BZ}}\!\frac{%
d^{2}q_{\perp }}{(2\pi )^{2}}\int_{0}^{\Lambda }\!\frac{dq_{z}}{2\pi }\frac{%
(u-1)^{2}}{u}\frac{1-e^{-uv_F^Bq_{z}\tau }}{q_{z}},  \nonumber\\ 
u^{2} &\equiv &1+(\kappa ^{B})^{2}\sum_{\bf G}({\bf Q}+{\bf G})^{-2},
\text{ }{\bf Q}=({\bf q}_{\perp },q_{z}).  \nonumber
\end{eqnarray}
\end{mathletters}
Here the $q_{\perp }$-integration goes over the Brillouin zone (BZ) of the
von Neumann lattice, ${\bf G}$ are the reciprocal lattice vectors,
$\Lambda $ is the cutoff, and $uv_F^B$ has the meaning of {\bf a} plasmon velocity. 
Typical $q_{z}$ in $f(\tau)$ are determined by the low-energy scale of the problem
($\rm{max}\{T,eV\}$) and thus small, which allows one to neglect 
$q_z$ in function $u$. 
The DOS reduces to the form
(\ref{nubulk}) with an exponent
\beq
\alpha=\pi\ell _{B}^{2}\int_{{\rm BZ}}\frac{
d^{2}q_{\perp }}{(2\pi )^{2}}\frac{
(u-1)^{2}}{u}.
\label{expbulk1}\eeq
For $\kappa^{B}\ell _{B}\!\ll\!1$, the sum over
the reciprocal vectors in (\ref{expbulk1}) is dominated
by the ${\bf G}\!=\!0$ term so that  
$u^{2}\!\approx\!1\!+\!(\kappa ^{B}/q_{\perp })^{2}$. 
The main contribution to the integral over $q_{\perp }$
in (\ref{expbulk1}) comes from the region $q_{\perp }\!\simeq\!
\kappa^{B}\!\ll\!1/\ell_{B}\!\simeq\!G$, thus the integration
can be extended over the entire $q$-space. The bulk DOS behaves
as $|\vareps-\vareps^B_F|^{\alpha}$, where $\alpha$ is
given in Eq.~(\ref{alphas}).

Note that in the main contribution to $\alpha$ [see
Eq.~(\ref{expbulk1})], a typical value of $u- 1$ is of the order of
unity. This indicates a non-perturbative nature of the result for
$\alpha$.  The lowest-order perturbation theory, applied to the bulk
case, would have given $\alpha\propto e^4$, whereas the correct result
is $\alpha\propto e^2$.

Tunneling into the end of a semi-infinite sample can be treated by imposing
the boundary condition on the current operator:
$j_{\bf R}\!=\!(1/i\sqrt\pi)\partial_\tau\phi_{\bf R}\!=\!(1/\sqrt\pi)
\partial_x\theta_{\bf R}\!=\!0$ for $z\!=\!0$.
This translates into the boundary conditions for the propagators:
${\cal P}^\rho_{\bf R}(z,z',\tau)\!=\!\langle\rho_{\bf R}(z,\tau)
\rho_{\bf R}(z',0)-\rho^2_{\bf R}(z,0)\rangle$, where
$\rho\!=\!\phi,\theta$. ${\cal P}^\rho_{\bf R}$
can be constructed from the propagators
for the infinite medium, $P^{\rho}_{\bf R}$,
by using the method of images. Evidently, 
${\cal P}^{\phi}_{\bf R}\!=\!0$ at the boundary whereas
${\cal P}^{\theta}_{\bf R}(0,0,\tau)\!=\!\pi^{-1}\int dq_z
P^\theta_{\bf R}(q_z,\tau)$.
The DOS at the boundary is again of form (\ref{nubulk})
but with an exponent
\beq
\alpha_s=2\pi\ell _{B}^{2}\int_{{\rm BZ}}\frac{%
d^{2}q_{\perp }}{(2\pi )^{2}}
(u-1). \label{fbound}
\eeq
In contrast to the bulk case, the 
logarithmically divergent integral over $q_{\perp}$ in (\ref{fbound})
is cut off at the BZ boundary, {\it i.e.}, for
$q_{\perp}\simeq 1/\ell_B$. Therefore, the condition of asymptotic orthogonality of CS is 
satisfied only with the logarithmic accuracy. The result for $\alpha_s$ is given
by Eq.~(\ref{alphas}).

In the strong-coupling case ($\alpha\gtrsim 1$), typical
$q_{\perp}\simeq 1/\ell_B$, which means that the approach based on CS
becomes invalid quantitatively. However, one can still estimate
tunneling exponents (both for the ``bulk'' and ``edge tunneling) as
$\tilde{\alpha}=C\sqrt{e^2/v_F^B}$, where the numerical constant $C\simeq 1$
cannot be determined by this method.

We are grateful to I. L. Aleiner, A. I. Larkin, and Z.\ Te\v{s}anovi\'c for
interesting discussions.
C. B. acknowledges the partial support of the project
COFIN98-MURST. The work at the University of Florida
was supported by the NHMFL In-House Research Program, 
NSF (DMR-970338) and Research
Corporation (RI0082). L. I. G. acknowledges the support
from NSF DMR-9731756 and DMR-9812340. 

\end{multicols}
\end{document}